\documentclass[twocolumn,showpacs,pra,tightenlines,eqsecnum,reprint]{revtex4-1}
\usepackage{amsmath}
\usepackage{amssymb}
\usepackage{graphicx}
\usepackage{bm}
\usepackage{upgreek}

\newcommand{\be}{\begin{equation}}
\newcommand{\ee}{\end{equation}}

\begin{document}

\title{Detection of the Abraham force with a succession of short optical pulses}

\author{Iver Brevik}
\author{Simen {\AA}. Ellingsen}
\affiliation{Department of Energy and Process Engineering, Norwegian
University of Science and Technology, N-7491 Trondheim, Norway}

\date{\today}

Revised version, July  2012

\begin{abstract}

For over a century, two rival descriptions of electromagnetic field momentum in matter have co-existed, due to Abraham and Minkowski, respectively.
We propose a set-up for measuring the difference between Abraham's and Minkowski's predictions in optics. To wit, a set-up is proposed in which the transient ``Abraham force'', a consequence of the Abraham energy-momentum tensor of 1909 may be measured directly. We show that when a train of short laser pulses 
is
sent through a fiber wound up on a cylindrical drum, the Abraham theory 
predicts
a torque which, by inserting realistic parameters, is found to be detectable. Indeed, the same torque when calculated with the Minkowski tensor takes the opposite sign. Numerical estimates show that with a typical torsion pendulum set-up
and standard laser parameters,
the angular deflection is in the order of $10^{-3}$ rad, which is easily measurable and even visible to the naked eye. Although its prediction is a century old, the Abraham force has proven experimentally elusive, and to our knowledge no 
macroscopic
experimental demonstration of the difference between the predictions of the two mentioned energy-momentum tensors exists at optical frequencies.
\end{abstract}


\maketitle

\section{Introduction}

It is remarkable-- and to the present authors rather surprising -- to see how great an 
attention
is presently 
given
to the classic energy-momentum problem in electrodynamics. The problem at hand is 
often dubbed
the Abraham-Minkowski problem: the rival candidates for an energy--momentum tensor for electromagnetic (EM) fields in media proposed by Abraham\cite{abraham09} and Minkowski \cite{minkowski10}, each of which may be convincingly argued for, do not appear to always predict the same physics. To wit, Abraham's tensor 
predicts
an additional force acting on dielectric bodies subjected to transient EM fields. This enigma, remembered to be considered a somewhat old-fashioned branch of physics back in the 1960's and 1970's, has 
been requickened to
become a modern topic again in 2012, as demonstrated by popular highlights \cite{volpe09} a series of recent experiments on its quantum photonics analog \cite{campbell05,hinds09,rikken11,wang11}, reported theoretical resolutions \cite{obukhov03,barnett10,mansuripur10,liu11, kemp11} and reviews \cite{loudon04,baxter10,barnett10b,milonni10}.

For all their virtues, purely theoretical treatments of the 
problem
cannot be expected to provide a simple answer to the question of which EM tensor is the more appropriate in general, since one invariably faces the fact that the electromagnetic energy-momentum tensor describes only a part (the EM field) of a coupled system including the material medium \cite{moller72,saldanha10}. Conservation laws concern the whole system, not its consituents, and a freedom of choice exists in the book-keeping of energy and momentum.

As we recently argued in a different context \cite{brevik11}, the choice of EM energy--momentum tensor is as much a question of practical convenience as one of correctness, a question which is most directly settled by asking which formalism can most easily describe observed effects. Experiments in which the Abraham force may be detected directly would be the ideal candidate. Of this category, the Walker--Lahoz--Walker experiment \cite{walker75,walker75b} remains the only one to our knowledge, making use of slowly varying, high amplitude electric fields. Analogous macroscopic experiments in the optical regime, most relevant to real-life applications, are still lacking. The effect observed in the experiment of She \emph{et al.} \cite{she08} can be described without reference to photon momentum \cite{brevik10}.

The set-up we propose herein bears some resemblance to that used by She's group, and is an optical analogy of the Walker--Lahoz--Walker experiment. The Abraham force, as well as the standard optical gradient force are transferred to a macroscopic cylinder as a torque, facilitating direct observation of the Abraham force. In fact, the angular deflection of the torsional balance has opposite sign whether Abraham's term is included in the force balance or not, rendering the observation simpler. We show that with realistic numbers for a laboratory set-up, the resulting deflections should 
conservatively
be in the order of 
$10^{-3}$ radians, which is readily measurable.
The system thus has several advantages over a previous suggestion involving whispering gallery modes in an optical resonator \cite{brevik10c}.

\section{Model: a long optical fiber}

Consider first the following model: a long dielectric circularly cylindrical rod  (fiber) of  length $L$ and radius $a$ is oriented along the $x$ axis. The refractive index in the material is the constant $n$, assumed to be  real. The rod is illuminated by a short laser pulse of total energy $\mathcal H_0$, duration $\tau$, frequency $\omega$, and original vacuum length $l_0= c\tau$; subscript   zero henceforth referring to  vacuum quantities. We assume for simplicity that $l_0 \ll L$ (this is easily generalized, but we 
endeavor
here to keep the formalism simple).  If the wave is just broad enough to fill out the cross section $A=\pi a^2$ of the rod,  we have   ${\mathcal H_0}=\varepsilon_0 E_0^2Al_0$, where $E_0$ is the rms value of the incident electric field.   For simplicity we assume that the ends are coated with antireflection films of refractive index $\sqrt n$, so that there is no reflected wave. Thus the energy in the medium, $\mathcal{H}=\varepsilon_0  n^2E^2Al$,   is the same as the incident energy $\mathcal H_0$. (Note that  $l_0=nl$, and that the continuity of Poynting's vector across the entrance region leads to the relationship $E_0^2=nE^2$.)

The general electromagnetic force density $\bf f$ in an isotropic nonmagnetic medium can be written as a sum of three physically distinct contributions (cf., for instance, Refs.~\cite{stratton41,brevik79}),
\be\label{1}
   {\bf f}={\bf f}^\mathrm{AM}+{\bf f}^\mathrm{ES} + {\bf f}^\mathrm{A}.
\ee
We will use SI units in the following.

The middle term in this expression,
${\bf f}^\mathrm{ES}$,
is the electrostriction term ($\rho$ denotes the material density). It is of importance in cases where knowledge about the distribution of pressure in a dielectric medium is needed (this point has recent been discussed, for instance,  in  Ref.~\cite{ellingsen11,ellingsen12}). However, as far as the total force on a body is concerned, the electrostriction term  does not contribute. We will therefore omit it in the following.

The first term in Eq.~(\ref{1}) is the force that acts in regions where $n$ varies, typically at the surfaces. It is common for the Abraham and Minkowski energy-momentum tensors, and may be called the Abraham--Minkowski force,
\begin{equation}
  {\bf f}^{\rm AM}=-\frac{1}{2}\varepsilon_0E^2 {\boldsymbol \nabla} n^2. \label{2}
\end{equation}
Its action on the left surface of the rod, at $x=0$, is to produce a surface pressure $\sigma_x$, directed to the left, from the medium side of the interface towards vacuum. It can be found by integrating the diagonal component of Maxwell's  stress tensor across the front surface,
\begin{equation}
  \sigma_x=\int_{0-}^{0+}T_{xx}dx= -\varepsilon_0 n(n-1)E^2, \label{3}
\end{equation}
$E$ being the field in the medium. The  surface impulse $G_{\rm surf}^{\rm AM}$  imparted to the front surface is thus
\begin{equation}
  G_{\rm surf}^{\rm AM}=\sigma_xA\tau=-(n-1)\frac{\cal H}{c}. \label{4}
\end{equation}
The transit time through the medium can with sufficient accuracy be put equal to $nL/c$. The impulse given to the exit surface is equal and opposite to the expression (\ref{4}), and the net momentum after the pulse has left is zero.

Consider next the last term in Eq.~(\ref{1}), the Abraham term,
\begin{equation}
 {\bf f}^{\rm A}=\frac{n^2-1}{c^2}\frac{\partial}{\partial t}(\bf E\times H). \label{5}
\end{equation}
It gives rise to an accompanying mechanical momentum $G_{\rm mech}^{\rm A}$ in the medium, equal to
\begin{equation}
 G_{\rm mech}^{\rm A}=\frac{n^2-1}{n}\frac{\cal H}{c}. \label{6}
\end{equation}
In the period when the pulse is contained in the medium, the pulse-induced total momentum is  thus
\begin{equation}
 G_{\rm surf}^{\rm AM}+G_{\rm mech}^{\rm A}=\frac{n-1}{n}\frac{\cal H}{c}. \label{7}
\end{equation}
Note already that the net impulse imparted including and excluding the Abraham term takes opposite signs.

Assume for simplicity that the rod is rigid, with  mass is $M=\rho AL$. As the expression (\ref{7}) must equal $Mv^{\rm A}$ where  $v^{\rm A}$ is the rod's velocity, the Abraham displacement $\Delta x^{\rm A}=v^{\rm A}Ln/c$ is
\begin{equation}
 \Delta x^{\rm A}=\frac{n-1}{\rho A}\frac{\cal{H}}{c^2}. \label{8}
\end{equation}
This displacement is formally independent of $L$ (although the above restriction $l\ll L$ has to be observed).

The expression (\ref{5}) is a small force on a macroscopic scale. In order to maximize its magnitude one would want high laser energy, high refractive index, and low mass per unit length $M/L$. Let us assume that a power of $P=1$ kW can be transmitted through the fiber (absorption and heat effects neglected). For illustration, take the duration of the pulse to be $\tau=10^{-8}$ s. With $n=3$  the length of the pulse is thus  $l_0 =3$ m  in vacuum and $l=1$ m in the medium. The pulse energy is ${\cal{H}}= P\tau=10^{-5}$ J. Taking the radius to be $a=10~\mu$m, and taking $\rho$=2000 kg/m$^3$, we  get
 \begin{equation}
 \Delta x^{\rm A}=3.5\times 10^{-16}~\rm m. \label{9}
 \end{equation}
 This is in practice unmeasurable. However -- and this is our main point -- one can make the effect much stronger by using a high-frequency {\it repetition} of pulses.

\section{Realistic manifestation and numerical results}

\begin{figure}[tb]
  \includegraphics[width=2.5in]{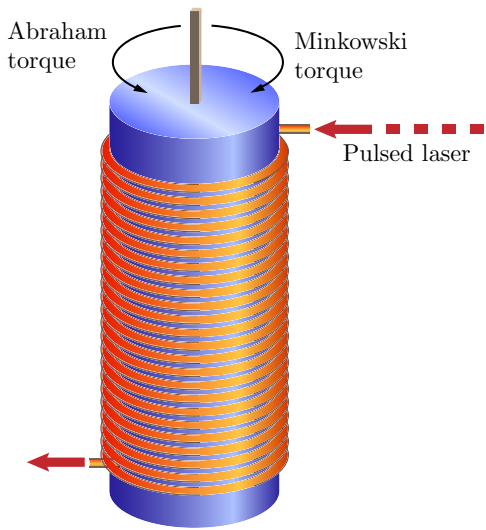}
  \caption{The proposed geometry: an optical fiber wound around a cylinder made of a light material. A train of optical pulses produces a torque, different for Abraham and Minkowski theory.}
\end{figure}

 The idealized picture of a rod as sketched above, micrometers thick and 
 meters long, is not practically useful. Let us instead assume that the fiber is flexible, and wound up on a low-mass cylindrical drum of radius $R \gg a$ and height   $H$. The system is hanging vertically in the gravitational field, suspended by a thin wire of known, small torsion constant $\kappa$. Let us assume for definiteness
 \begin{equation}
   R=10~\rm{cm}, \quad H=10~\rm cm. \label{10}
 \end{equation}
 If the cylinder is made of a dilute material, such as polystyrene whose density is about 100 kg/m$^3$, the cylinder mass $M_{\rm cyl}$ and its moment of inertia $I_{\rm cyl}=\frac{1}{2}M_{\rm cyl}R^2$ about the $z$ axis become approximately
 \begin{equation}
 M_{\rm cyl} = 31 ~\rm{g}, \quad I_{\rm cyl} = 1.6\times 10^{-3}~\rm kg~m^2. \label{11}
 \end{equation}
 The cylinder could of course be made hollow to further decrease its mass and moment of inertia.

 \subsection{Abraham theory}

 Winding a fiber of diameter $2a=20~\mu$m around the cylinder, there is room for $H/2a$ circuits in a single layer of windings. Without multiple layers of windings, this already corresponds to $5000$ turns of circumference $2\pi R$, i.e., $L\approx 3$km. The single-pulse momentum (\ref{7}), when multiplied with $R$ and with the pulse repetition frequency  $f_{\rm rep}$, is the angular momentum supplied to the cylinder per second. This is the same as the torque $N_z^{\rm A}$ about the $z$ axis. The cylinder will slowly turn abound the $z$ axis, until mechanical equilibrium is restored. The maximum angular deflection, called $\phi_{\rm max}^{\rm A}$, follows as
 \begin{equation}
 \phi_{\rm max}^{\rm A}=\frac{1}{\kappa}N_z^{\rm A}=  \left(\frac{n-1}{n}\frac{\cal{H}}{c}\right) \frac{R}{\kappa}f_{\rm rep}. \label{13}
 \end{equation}
 This expression is independent of $L$. The usefulness of having a large value of $L$ is that this easily ensures the pulses to be spatially separated  in the medium. The main parameter in the above expression, in order to obtain a large value of $\phi_{\rm max}$, is seen to be the product ${\cal{H}} f_{\rm rep}$.

 In order to make clear the main idea behind our model let us assume again that the medium is rigid (thus ignoring its real elasticity properties), and let us take it to be long and straight. During the time the first pulse is in the fiber, the fiber undergoes a displacement $\Delta x^{\rm A}$,  given by Eq.~(\ref{8}). A subsequent pulse, sufficiently delayed by a time interval $\Delta t$  not to  overlap with the first, provides another displacement $\Delta x^{\rm A}$. Assume  for definiteness that the pulse train contains $N$ pulses in all, and that they are all contained in the fiber at the same time.  Once the first pulse exits the fiber, the impulse from this pulse is canceled out, the action from the other $(N-1)$ pulses however remaining. The subsequent pulses act in the same way. The total displacement effect is additive, so that the total displacement is  equal to the 
 one-pulse displacement $\Delta x^{\rm A}$ multiplied by $N$. When the pulse train has left, the position of the fiber is thus changed, but 
 its residual momentum (assuming no absorption)
 is zero.
 It is possible to interpret the  force  as a transient effect, as any  pulse when contained in the fiber contains a separate momentum of its own. The force accumulates the momenta of the pulses contained in the fiber at any time.

Returning to the cylinder geometry, when the number of repeated pulses is high the deflection can be enhanced considerably.  Repetition rates in the multi gigaherz regime have been experimentally achieved for a long time \cite{tai86,longhi00, vlachos00}. In these experiments the average optical power is typically in the range of a few tens of mW, although high output power is not the objective. Let us assume the same value for $\tau$ as above, but moderate the input value of $P$ to make it more realistic:
\begin{equation}
P= 1~\rm{W}, \quad \tau=10^{-8}~\rm{s}, \quad {\cal{H}}=10^{-8}~\rm J.
\end{equation}
To calculate $\phi_{\rm max}$ we have moreover to
 estimate a value for the torsion constant $\kappa$. In extreme cases, such as when dealing with torsion experiments testing the equivalence principle \cite{hou03,schlamminger08}, the torsion constant has been reported as low as about $10^{-9}$ Nm/rad. Let us adopt a somewhat larger value here,
 \begin{equation}
 \kappa \sim 10^{-8}~\rm Nm/rad. \label{15}
 \end{equation}
 Insertion into Eq.~(\ref{13}) now gives
 \begin{equation}
 \phi_{\rm max}^{\rm A} =2.2\times 10^{-10}f_{\rm rep}.       \label{A}
 \end{equation}
 Let us for definiteness assume that the separation between the centers of each pulse (each of length $l=1~$m) is 10 m. Then there are $L/10=300$ pulses in the fiber at the same time. To propagate 10 m in the fiber, light needs $10^{-7}$ s (the entrance time $\tau$, as mentioned, is only one tenth of this). It thus seems reasonable to adopt as repetition frequency
 \begin{equation}
 f_{\rm rep}=10~\rm MHz.
 \label{B}
 \end{equation}
Then Eq.~(\ref{A})  leads to the estimate
\begin{equation}
\phi_{\rm max}^{\rm A} = 2.2\times 10^{-3} \,\rm rad. \label{C}
\end{equation}
A deflection of this magnitude should be easily measurable.  The expression is positive, meaning that the cylinder turns in the 
direction of light propagation.

It should be emphasized that the above argument rests upon the assumption of additivity: the effective force on the fiber is found by multiplying the impulse transferred from one single pulse by the number of pulses transmitted through the fiber {\it per second}. It is not necessary that each pulse enters and exits the fiber before the next pulse enters. The important point is merely that the pulses are separated from each other in the fiber.

A comment on sources of corrections is warranted.
Take again the model of a long straight rod and assume that the antireflection films on the ends are not perfect, but that there is an effective  
reflectivity coefficient
$\cal R$ at each end. Assuming ${\cal{R}} \ll 1$ we may assume that the fields in the medium are practically the same as in the ideal case considered above.  Since a fraction $\cal R$ of the incident energy is reflected at the front surface we estimate that, during the entrance period of each pulse, an extra positive impulse equal to $\cal R$ times the magnitude $|G_{\rm surf}^{\rm AM}|$ of the surface impulse (\ref{4}) is imparted to the rod,
\begin{equation}
  {\cal{R}} |G_{\rm surf}^{\rm AM}| = {\cal{R}}(n-1)\frac{\cal H}{c}. \label{D}
\end{equation}
During the exit time of the same pulse the same extra amount is imparted to the rod, also then in the forward direction. Adding this to Eq.~(\ref{7}) we have
\begin{equation}
  G_{\rm surf}^{\rm AM}+G_{\rm mech}^{\rm A} \rightarrow (1+2n{\cal R})\frac{n-1}{n}\frac{\cal H}{c}, \label{E}
\end{equation}
implying that the relative correction to the total momentum is simply $2n\cal R$. Thus the required accuracy of the reflection coefficient is not too demanding; with $n=3$ as assumed above we see that ${\cal R} = 0.005$ is sufficient to give an accuracy of about 3\% in the momentum. 
In the same way the small but nonzero absorption of momentum from the propagating light by the fiber must be accounted for as a correction.

 \subsection{Minkowski theory}

 The above theory concerns the Abraham theory. Let us consider Minkowski's theory in which the Abraham term is excluded. The Minkowski energy-momentum tensor has in general several advantages, not least so in optics: it is divergence-free in a homogeneous medium without external charges implying  that the four components of energy and momentum make up a four-vector \cite{moller72}, and   it moreover adjusts itself very nicely to a canonical treatment \cite{brevik70,barnett10}, in spite of its space-like character (the field energy can be negative in some inertial systems).

 In our case, the  difference from the Abraham case 
 appears
 in the omission of the  last term in Eq.~(\ref{1}), and similarly in the omission of the term $G_{\rm mech}^{\rm A}$ in Eq.~(\ref{7}). The contribution to the mechanical momentum in the rod comes entirely from the surface forces. The linear displacement $\Delta x^{\rm M}$ predicted in the Minkowski theory thus becomes
 \begin{equation}
 \Delta x^{\rm M}=-\frac{n(n-1)}{\rho a^2}\frac{\cal H}{c^2}, \label{16}
 \end{equation}
 instead of Eq.~(\ref{8}) (note that the energy $\cal H$ is the same). Similarly we get
 \begin{equation}
 \phi_{\rm max}^{\rm M}=-(n-1)\frac{\cal H}{c}\frac{R}{\kappa}f_{\rm rep}. \label{17}
 \end{equation}
 That means,
 \begin{equation}
 \phi_{\rm max}^{\rm M}=-n\phi_{\rm max}^{\rm A}.
 \end{equation}
 With the numbers employed above, this amounts to
 \be
   \phi_{\rm max}^{\rm M}=-6.6 \times 10^{-3}~\rm rad.
 \ee

 The most important difference from Eqs.~(\ref{13}) and (\ref{C}) is the difference in {\it sign}. The cylinder is predicted to turn in the opposite direction from the case above. This point obviously serves to facilitate possible forthcoming experiments.

 There is of  course  ample room for choosing different values for the input parameters than those used in our numerical estimates above. However,  our main point has  been to demonstrate that the simple trick of using repetitive pulses should make it 
 possible to measure the difference between the Abraham and Minkowski predictions in optics in an experiment which is simple, at least 
 conceptually.
 It would be of definite interest to see such an experiment carried out in practice.

 \section{Conclusions}

 The two main contestants for the energy--momentum tensor in media are those due to Abraham and Minkowski, both presented more than a century ago, yet still the question of which one to choose attracts considerable attention. While several experiments have been carried out in recent years probing the question at an atomic and photonic scale, experiments in macroscopic electromagnetics have been scarce due to the Abraham force's small magnitude and difficulty of access.

 Here we have proposed a set-up which might make such a measurement possible, by which the difference between (na\"{\i}ve) Abraham and Minkowski predictions can be observed, perhaps even with the naked eye. While this certainly would not prove either tensor ``right'' (it is a fallacy to talk 
 of 
 correctness 
 --- rather it is a question of usefulness), it would provide a strong case for the usefulness of 
 the
 concept of the Abraham force in explanation or prediction of optical forces on matter interacting with transient electromagnetic fields.

\end{document}